\documentclass[preprint,nocite]{epl}
\usepackage{graphicx}

\title{Low-energy spin excitations in the molecular magnetic cluster $\chem V_{15}$}

\shorttitle{Low-energy spin excitations in $\chem V_{15}$}

\author{G.~Chaboussant\inst{1}\thanks{E-mail: \email{chabouss@iac.unibe.ch}}, R.~Basler\inst{1},
        A.~Sieber\inst{1}, S.T.~Ochsenbein\inst{1}, A.~Desmedt\inst{2},  R.E.~Lechner\inst{2},
        M.T.F.~Telling\inst{3}, P.~K\"{o}gerler\inst{4}, A.~M\"{u}ller\inst{5} \and H.-U.~G\"{u}del\inst{1}}

\shortauthor{G.~Chaboussant \etal}

\institute{
   \inst{1} Departement f\"{u}r Chemie und Biochemie, Universit\"{a}t Bern - CH-3000 Bern 9,
   Switzerland.\\
   \inst{2} BENSC, Hahn-Meitner Institut Berlin - D-14109 Berlin,
   Germany.\\
   \inst{3} ISIS Facility, Rutherford Appleton Laboratory - Didcot, Oxon 0X11 0QX,
   UK.\\
   \inst{4} Ames Laboratory, Iowa State University - Ames, IA50011,
   USA.\\
   \inst{5} Fakult\"{a}t f\"{u}r Chemie, Universit\"{a}t Bielefeld - 33501 Bielefeld,
   Germany.\\
}


\begin{document}

\maketitle

\begin{abstract}
We report an Inelastic Neutron Scattering (INS) study of the
fully deuterated molecular compound $\chem
K_{6}[V^{IV}_{15}As_{6}O_{42}] \cdot 9D_{2}O$ ($\chem V_{15}$).
Due to geometrical frustration, the essential physics at low
temperatures of the $\chem V_{15}$ cluster containing $15$ coupled
$\chem V^{4+}$ (S=1/2) is determined by three weakly coupled
spin-1/2 on a triangle. The INS spectra at low-energy allow us to
{\it directly} determine the effective exchange coupling $J_{0} =
0.211(2)$ meV within the triangle and the gap $2 \Delta = 0.035(2)
$ meV between the two spin-1/2 doublets of the ground state.
Results are discussed in terms of deviations from trigonal
symmetry and Dzyaloshinskii-Moriya (DM) interactions.
\end{abstract}

\section{Introduction}

The recent discovery of quantum tunnelling of the magnetisation
vector in magnetic cluster complexes with high-spin ground state
(GS), for example $\chem Mn_{12}$-acetate with $S=10$
\cite{Sessoli93b,Gatteschi94,Thomas96}, has triggered a renewed
interest in molecular magnets containing a small group of
metallic ions embedded in organic or inorganic ligands. In such
structures, each metallic cluster is magnetically isolated from
its neighbours, and this allows the study of the collective
behaviour and inherent quantum size effects within a finite
number of magnetic atoms. Such molecular Magnets with {\it
high-spin} GS prove to be extremely valuable candidates to study
phenomena, critical at the nanoscopic scale, such as slow
relaxation and quantum tunnelling of the magnetisation
\cite{Barbara00}, quantum coherence \cite{Luis00} and Berry
phases \cite{Wernsdorfer99}. Amongst the existing materials,
systems with {\it low-spin} GS like the polyoxovanadate $\chem
K_{6}[V_{15}As_{6}O_{42}]\cdot 9H_{2}O$, shortnamed $\chem
V_{15}$ hereafter, are of the greatest interest as they are shown
to display quantum coherence and slow relaxation of the
magnetisation despite their low-spin GS and the absence of any
magneto-crystalline energy barrier \cite{Chiorescu00,Chiorescu01}.

In this context, INS is an extremely valuable technique to
determine the key microscopic parameters controlling the
tunnelling mechanism in high-spin clusters such as $\chem
Mn_{12}$-acetate \cite{Mirebeau99} and $\chem Fe_{8}$
\cite{Caciuffo98}. In this Letter we report the first Inelastic
Neutron Scattering (INS) study of a fully deuterated $V_{15}$
polycrystalline powder. Frustration of the exchange interactions
plays a major role in its low-energy properties which can be
accurately described by three weakly coupled spin-1/2. INS allows
us to determine directly the effective exchange interaction
$J_{0} = 0.211(2)$ meV between the spins in the triangle and the
GS splitting $2 \Delta = 0.035(2)$ meV.

\section{Magnetic properties}

\begin{figure}
\onefigure[scale=0.6]{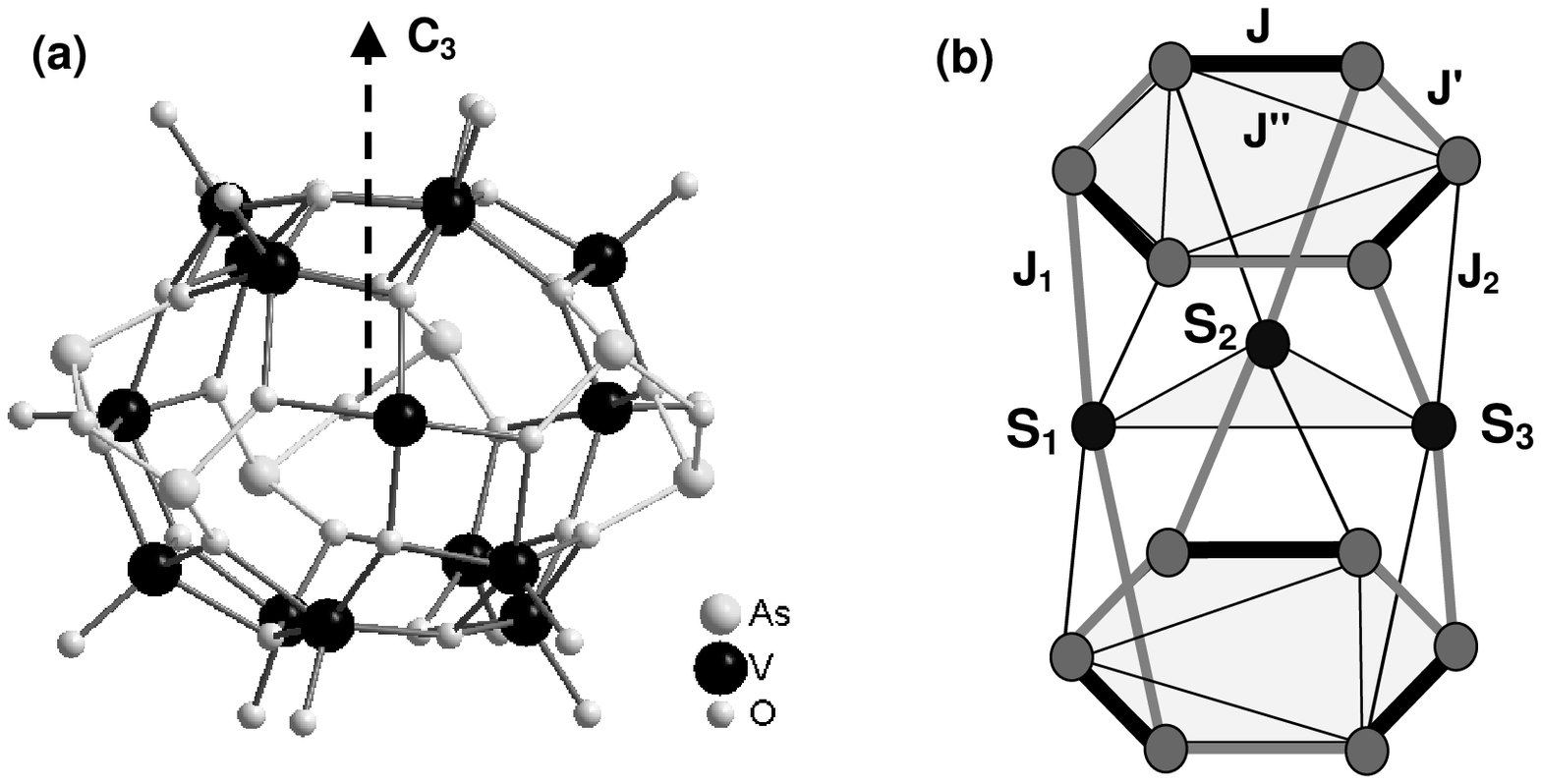}
%\centerline{\includegraphics[height=50mm,angle=0]{figure1}}
\caption{(a)~Structure of the cluster anion in $\chem
[V_{15}As_{6}O_{42}]^{6-}$ cluster. The 15 $\chem V^{4+}$ (dark
solid spheres) form two outer distorted hexagonal layers and an
inner triangular layer. (b)~Schematic representation of the AFM
exchange couplings between $\chem V^{4+}$ ions according to the
type of oxygen bridges involved in the exchange paths (distance
and angle). The magnetic Hamiltonian is defined as ${\cal H} =
\sum_{<i,j>} J_{ij}\bm{S}_{i}\bm{S}_{j}$ with $<i,j>$ pairs as
indicated. $\bm{S}_{1}$, $\bm{S}_{2}$ and $\bm{S}_{3}$ in the
triangle are loosely connected through the hexagons and
diarsenite bridges.} \label{fig:1}
\end{figure}

$\chem V_{15}$ is a polyoxovanadate molecular magnet made of 15
spins s=1/2 (all $\chem V^{4+}$ ions) \cite{Muller88}. The
symmetry is trigonal (space group $R\bar{3}c$, $a=14.029$ \AA\,
$\alpha = 79.26^{o}$). The cluster has a local $C_{3}$ symmetry
with the $V-V$ distances varying between $2.87$ \AA\ and $3.05$
\AA\ ; the intermolecular dipolar couplings are very small (few
mK) and will be ignored in most situations. As first discussed in
Refs. \cite{Gatteschi91,Barra92}, the $\chem V^{4+}$ (S=1/2) ions
are arranged in three layers as shown in fig.~\ref{fig:1}a and
are antiferromagnetically (AFM) coupled to their neighbours via
oxo-bridges. Each of the two hexagons in fig.~\ref{fig:1}b
contains three pairs of strongly coupled spins through oxo-bridge
pathways between the $\chem (VO_{5})$ square pyramids. Three
types of couplings, shown in fig.~\ref{fig:1}b, are of importance
for the magnetic properties: Nearest neighbour couplings $J
\approx 70$ meV and $J' \approx 13$ meV and next-nearest
neighbour couplings $J" \approx 26$ meV. In addition, each spin
of the central triangle is coupled to the hexagons via $J_{1}
\approx J'$ and $J_{2} \approx J"$. Crucially, the three spins
S=1/2 in the triangle are {\it not directly} coupled to each
other by oxygen bridges like in the hexagons. Two exchange
pathways between the triangle spins should be considered: One
exchange pathway occurs via the upper and lower hexagons with
$\chem d(V-O-V_{hex}-O-V) \approx 10 $ \AA\ while the other is
mediated by diarsenite bridges with $\chem d(V-O-As-O-As-O-V)
\approx 10 $ \AA\ . The latter coupling can be compared to an
analogous complex, $\chem V_{12}$ \cite{Basler02}, where the
coupling through arsenite bridges is $J \approx 0.7-0.8$ meV with
$\chem d(V-O-As-O-V) \approx 5.3 $ \AA\ . The $\chem V-V$ distance
being much larger in the title compound, one expects the
couplings through the diarsenite bridges to be much smaller than
$0.7$ meV in $V_{15}$.

\begin{figure}
\onefigure[scale=0.6]{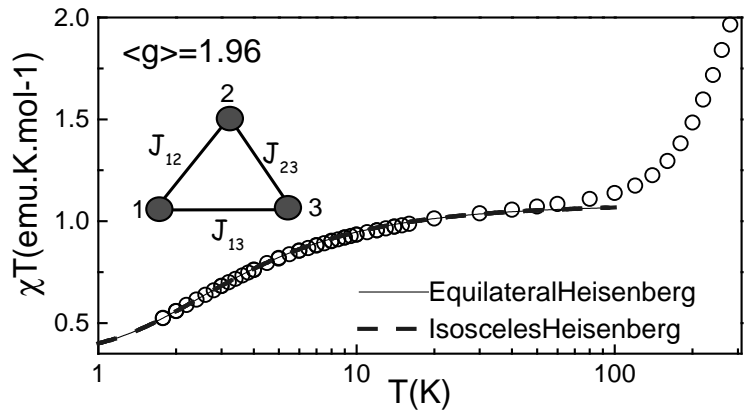}
%\centerline{\includegraphics[height=50mm,angle=0]{figure2}}
\caption{DC SQUID magnetic susceptibility obtained on a $160$ mg
powder sample of deuterated $\chem V_{15}$ at an applied external
field $H=0.1$ T. The powder averaged Land\'e g-factor is set to
$1.96$ from EPR data: $g_{a}=g_{b}=1.95$, $g_{c}=1.98$
\cite{Gatteschi91,Barra92}. Solid and dashed lines are calculated
curves from eq.\ref{eq:Heis} for an equilateral triangle ($J_{12}
= J_{23} = J_{13} = 0.211$ meV) and an isosceles triangle
($J_{12} = J_{23} = 0.223$ meV and $J_{12} = 0.188$ meV),
respectively (see text).} \label{fig:2}
\end{figure}

Figure.~\ref{fig:2} shows the DC susceptibility represented as
$\chi T$ versus $T$ for the fully deuterated version of $V_{15}$.
No difference in the $\chi T$ could be observed with the
undeuterated compound \cite{Gatteschi91}. The essential
information from the susceptibility can be summarized as follows:
At $T=300$ K the $\chi T$ product does not saturate as expected
for a truly paramagnetic regime thereby indicating that the spins
are still strongly correlated. Below $100$ K, the $\chi T$ curve
exhibits a plateau at around $1.1 \un{emu.K.mol^{-1}}$, a value
consistent with three uncoupled s=1/2 spins. At lower
temperature, a further drop of $\chi T$ suggests a S=1/2 GS. In
this regime, the dimers in the hexagons are firmly in a singlet
state ($T < 0.1J$) and the cluster magnetic properties can be
accounted for by three spin-1/2 coupled by weak {\it effective}
antiferromagnetic exchange interactions within the central
triangle. An obvious starting point is then the $S=1/2$ AFM
Heisenberg model on a triangle:
\begin{equation}
{\cal H}_{0} = J_{12} \bm{S}_{1}\bm{S}_{2} + J_{23}
\bm{S}_{2}\bm{S}_{3} + J_{13} \bm{S}_{1}\bm{S}_{3} \; ,
\label{eq:Heis}
\end{equation}
where $\bm{S}_{1}$, $\bm{S}_{2}$, $\bm{S}_{3}$ denote the spin
operators on sites $1$,$2$ and $3$, respectively (see
fig.\ref{fig:1}b) and $J_{ij}$ is the Heisenberg exchange
parameter between spin $i$ and $j$. We first assume the
equilateral triangle case: $J_{ij} = J_{0}$ for all $(ij)$ pairs.
For an antiferromagnetic coupling ($J_{0}>0$), the GS consists of
two degenerate S=1/2 Kramers doublets separated from the $S=3/2$
excited state by an energy $3J_{0}/2$. The wave functions of the
two degenerate S=1/2 Kramers doublets are given by
$\Psi^{\pm}_{0} = \mid 0,\frac{1}{2},\frac{1}{2},\pm
\frac{1}{2}>$ and $\Psi^{\pm}_{1} = \mid
1,\frac{1}{2},\frac{1}{2},\pm \frac{1}{2}>$ in the basis $\mid
S_{12},S_{3},S,M_{S}>$ or any linear combination of them.

As a function of magnetic field the $M_{S}=-3/2$ component of the
$S=3/2$ spin state crosses the $M_{S}=-1/2$ level of the ground
state at a critical field $H_{0} = 3J_{0}/2g\mu_{B}$. Recently,
Chiorescu {\it et al.} observed a jump in the low-temperature
magnetisation at $H_{0} \approx 2.8$ T \cite{Chiorescu01} which
gave the first evaluation of the effective coupling: $J_{0} =
0.211$ meV. The calculated susceptibility using the Heisenberg
model of eq.~\ref{eq:Heis} and $J_{0} = 0.211$ meV matches very
well our experimental $\chi T $ versus $T$ behaviour between $2$ K
and $40$ K as shown in fig.~\ref{fig:2}. However this simple
model must be refined according to several experimental results.
Firstly, the magnetisation curve at low temperature exhibits an
anomalous broadening of the jump at $H_{0}$ \cite{Chiorescu01}
which could not be explained by the equilateral triangular
Heisenberg model and, more importantly, magnetisation relaxation
experiments \cite{Chiorescu00} suggest the presence of a {\it
tunnel splitting} $2\Delta$ of the order of $9\un{\mu eV}$ within
the $S=1/2$ GS. This was ascribed to antisymmetric
Dzyaloshinskii-Moriya (DM) terms \cite{Dobrovitski00}. In order
to gain more insight, we have undertaken an INS study to directly
probe the low-energy levels of this material.

\section{Neutron scattering experiment}

\begin{figure}
\onefigure[scale=0.6]{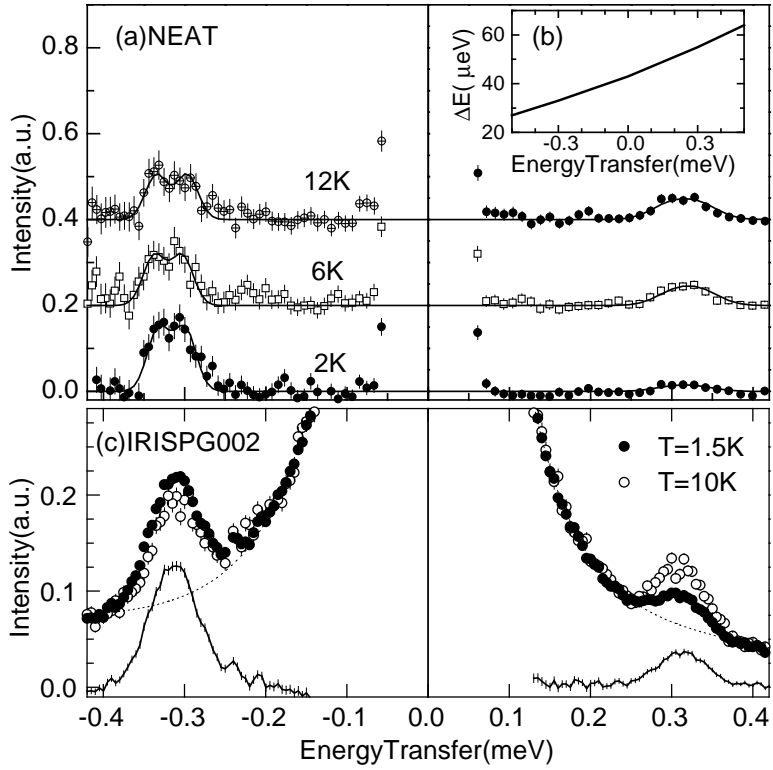}
%\centerline{\includegraphics[height=100mm,angle=0]{figure3}}
\caption{(a)~INS spectra of fully deuterated $\chem V_{15}$
obtained on NEAT (HMI) at $\lambda_{i} = 8.0 $ \AA\ with detector
angles $ 23^{o} \leq 2\theta \leq 71^{o}$. Data were collected at
several temperatures and corrected for the background (empty
sample holder) and detector efficiency (Vanadium metal
reference). The solid lines are fits to the data as explained in
the text. (b)~Instrumental resolution $\Delta E$ as a function of
the energy transfer $\hbar \omega$ calculated for NEAT. The
calculation of the energy dependence of $\Delta E$ is performed
according to NEAT specifications \cite{Lechner91} and rescaled by
hand to match the measured elastic energy resolution $\Delta
E_{0}$. (c)~INS spectra at $1.5$ K and $10$ K obtained on IRIS
(ISIS) with $\lambda_{f} = 6.66 $ \AA\ $^{-1}$ and for detector
angles $28.5^{o} \leq 2\theta \leq 128^{o}$. Data were corrected
similarly as for NEAT. The $T=1.5$ K Data after subtraction of
the thermal diffuse scattering is shown a solid lines.}
\label{fig:3}
\end{figure}

We used a $m \sim 4$ g polycrystalline sample of fully deuterated
$\chem V_{15}$ placed under Helium in a rectangular flat $\chem
Al$ slab ($3 \times 5$ cm) of $3$ mm thickness. The INS
experiments were performed on the Time-of-Flight (TOF)
spectrometer NEAT \cite{Lechner01} at the BENSC (Hahn-Meitner
Institute, Germany) using cold neutrons with an incident
wavelength of $\lambda_{i} = 8.0 $ \AA\ and at the ISIS Facility
(Rutherford Appleton Laboratory, United Kingdom) using the
backscattering TOF spectrometer IRIS with a final neutron
wavelength of $\lambda_{f} = 6.66 $ \AA\ and the pyrolytic
graphite (PG002) analyzer. On NEAT, the angle between the
incoming beam and the plane of the sample holder was $135^{o}$
and the experimental energy resolution $\Delta E_{0}$, measured
from a vanadium sample and averaged over all scattering angles,
was $43\un{\mu eV}$ for zero energy transfer. The energy
resolution $\Delta E$ depends on the energy transfer $\hbar
\omega$ \cite{Lechner91,Lechner84} as shown in fig.~\ref{fig:3}b.
Data reduction was carried out using the program INX, and the
fitting procedure included the energy dependence of the
instrumental resolution. On IRIS, the experimental resolution
varies from $17\un{\mu eV}$ at the elastic position to $\approx
28 \pm 2 \un{\mu eV}$ at $\hbar \omega \approx \pm 0.3$ meV. The
program ICON was used for data reduction.

The low-energy part of the scans obtained on NEAT is depicted in
fig.~\ref{fig:3}a at several temperatures with an average
momentum transfer $Q_{av} \approx 0.6 $ \AA\ $^{-1}$
\cite{comment1}. At $T=2$ K, the energy loss side (left-hand side)
shows a single broad peak centered at $\hbar \omega \approx
-0.315$ meV. As the temperature increases, a peak at the same
energy appears on the energy gain side. The instrumental
resolution at the inelastic peak positions is $32 \un{\mu eV}$
and $55 \un{\mu eV}$ in loss and gain, respectively, as shown in
fig.~\ref{fig:3}b. The observed peaks have weakly
temperature-dependent widths with FWHM values between $57$ and
$76 \un{\mu eV}$ on the loss side and between $82$ and $100
\un{\mu eV}$ on the gain side. This is about twice the
instrumental resolution, and we conclude that the peak contains
more than one transition. In fact, the peaks have a flat-top
shape in energy-loss suggesting that the observed broad peak is
made of two unresolved INS transitions. Figure.~\ref{fig:3}c
shows the low-energy part of the spectrum obtained on IRIS at
$1.5$ K and $10$ K with with an average momentum transfer $Q_{av}
\approx 1.1 $ \AA\ $^{-1}$ \cite{comment1}. The peaks at $\hbar
\omega = \pm 0.315$ meV correspond to those observed on NEAT but
are superimposed on the thermal diffuse scattering of the PG002
analyzer. Again, the peaks on either side of the elastic line are
broader than the instrumental resolution.

The mean energy of the peak corresponds to the energy expected
for the $S=1/2 \leftrightarrow S=3/2$ INS transition in the
equilateral Heisenberg model ($\hbar \omega_{0} = 3J_{0}/2 =
0.315$ meV) with $J_{0} = 0.211$ meV inferred from the
magnetisation jump at $H \approx 2.8$ T. To account for the
broader-than-resolution peak, we analysed the data as follows:
The broad peak centered at $0.315$ meV was fitted using two
Gaussians at positions $\hbar \omega^{\pm}_{0} = 3J_{0}/2 \pm
\Delta$. The widths are kept fixed at the instrumental resolution
values. Only $J_{0}$, $\Delta$ and an overall
temperature-dependent intensity scaling factor, $I_{T}$, are
allowed to float. In a first approximation, $I_{T}$ is set equal
for both transitions as no significant differences can be
observed in their intensities within the experimental accuracy.
The best fit is obtained for $J_{0} = 0.211 \pm 0.002$ meV and
$2\Delta = 0.035 \pm 0.002$ meV. The results are shown as solid
lines in fig.~\ref{fig:3}a.

\begin{figure}
\onefigure[scale=0.6]{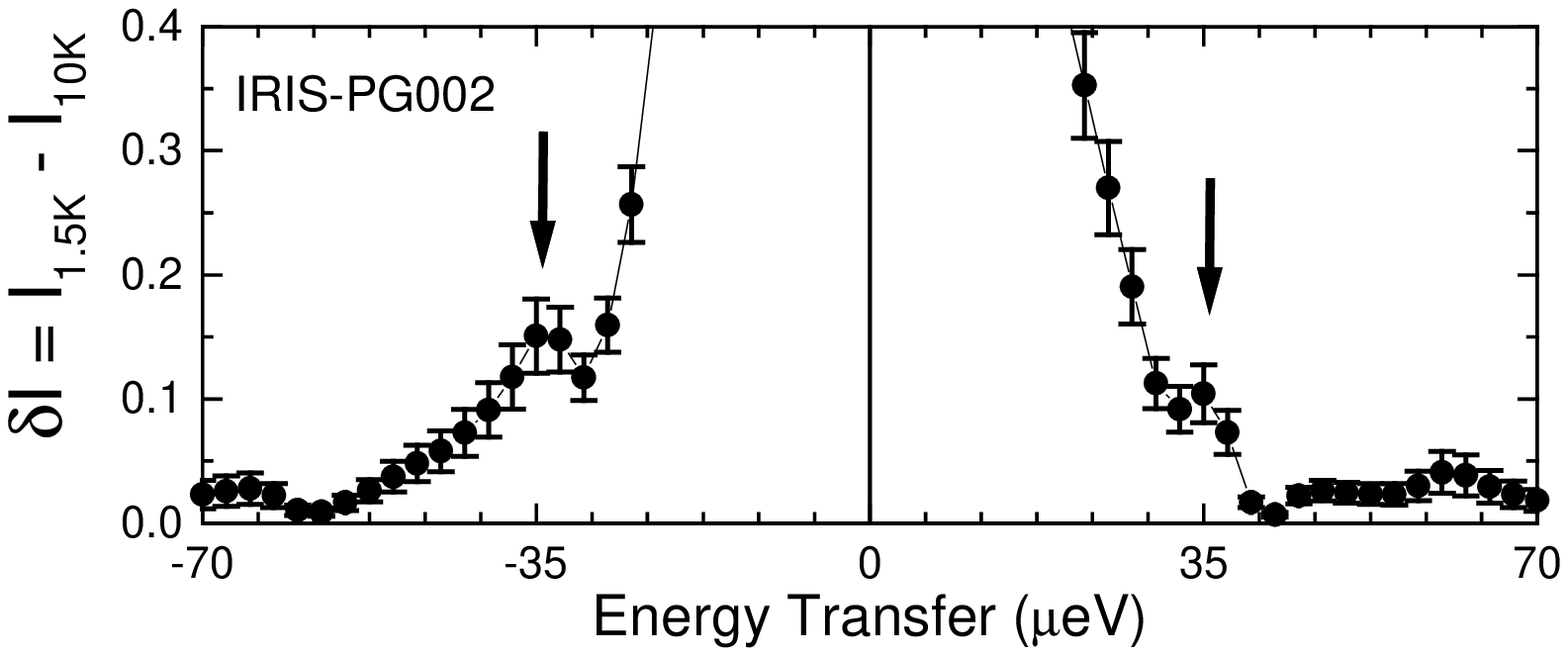}
%\centerline{\includegraphics[height=45mm,angle=0]{figure4}}
\caption{Difference between the $1.5$ K and the $10$ K IRIS data.
The arrows indicate the positive peaks discussed in the text.}
\label{fig:4}
\end{figure}

From these transitions alone, it is not possible to assess whether
the splitting originates from the S=1/2 GS doublets or from the
S=3/2 excited state. In both cases, the thermal energy, even at
the lowest temperature ($2$ K), is much larger than $2\Delta
\approx 0.4$ K, and the change in the population factor between
$2$ K and higher temperatures is too small to be measured in the
{\it relative intensities} of the two transitions $\hbar
\omega^{\pm}_{0}$. However, by subtracting the $10$ K data from
the $1.5$ K data shown in fig.~\ref{fig:3}c, it is apparent that
two weak positive peaks occur at $\hbar \omega^{\pm}_{1} = \pm
0.035$ meV. Figure.~\ref{fig:4} shows the difference $\delta I =
I_{\ab{1.5K}}- I_{\ab{10K}}$ in the low-energy region $| \hbar
\omega | \leq 70 \un{\mu eV}$. These two peaks are symmetric
around the elastic line and their energies coincide {\it exactly}
with the value of the splitting inferred from the higher energy
peaks. Even though the accuracy of the subtracting procedure is
not good enough to analyze them quantitatively, the fact that the
transitions at $\hbar \omega^{\pm}_{1}$ have a larger intensity
at $1.5$ K than at $10$ K strongly suggests that the splitting is
within the two S=1/2 doublets and not within the S=3/2 excited
state \cite{comment2}.

\section{Analysis}

To rationalise the observed splitting $2\Delta$, the equilateral
triangular Heisenberg model must be refined. Two types of
processes can be put forward: (a) Antisymmetric
Dzyaloshinskii-Moriya (DM) interactions \cite{DM}, which occur as
a result of spin-orbit coupling in non-centrosymmetric binuclear
units (this is true for the edge sharing $\chem VO_{5}$ units in
$\chem V_{15}$) and (b) deviations from purely trigonal symmetry:
Removing the restriction $J_{12}=J_{13}=J_{23}$ in
eq.~\ref{eq:Heis} will release part of the exchange frustration
and lift the degeneracy of the S=1/2 ground state.

Chiorescu {\it et al} \cite{Chiorescu00} first invoked DM terms as
being responsible for the splitting in the ground state of
$V_{15}$. The DM Hamiltonian may be written as
\begin{equation}
{\cal H}_{\ab{DM}} = \sum_{ij} \bm{D}_{ij} . (\bm{S}_{i} \times
\bm{S}_{j}) \; , \label{eq:DM}
\end{equation}
where $\bm{D}_{ij}$ is the DM vector for the pair $(ij)$ between
$\bm{S}_{i}$ and $\bm{S}_{j}$ along the triangle. In pure
trigonally symmetric trinuclear systems, $\bm{D}_{ij}$ is
identical for each bond pair and points parallel to the unique
trigonal axis ($\bm{D}_{ij} = D\bm{z}$) possibly leading to a
non-collinear $120^{o}$-type spin arrangement in the $XY$ plane .
Including DM interactions in the total Hamiltonian, ${\cal H}_{1}
= {\cal H}_{0} + {\cal H}_{DM}$, generates a splitting $2
\Delta_{1} = \sqrt{3}D$ between the two $S=1/2$ Kramers doublets
of energy $E^{\pm} = \pm \Delta_{1}$, and the eigenfunctions are
now given by specific mixture of $\Psi^{\pm}_{0}$ and
$\Psi^{\pm}_{1}$ \cite{Rakitin81,Fainzil81}. The observed
splitting, $2\Delta = 0.035$ meV would then lead to $D \approx
0.02$ meV. The order of magnitude of $D$ can be theoretically
estimated from the Moriya expression \cite{DM}: $D \approx
(\Delta g /g_{0})J_{0}$ where $\Delta g$ is the deviation of the
average g-factor value, $g=1.96$, from the free-ion value
$g_{0}=2.00$. In our case, this would lead to a theoretical $D$
of $ \approx 4.2 \un{\mu eV}$, a value $4.7$ times smaller than
the value deduced from the experimentally determined $2\Delta$.
This discrepancy is indicative that the GS splitting is most
probably not solely due to DM interactions. Next, we consider
deviations from trigonal symmetry. Elements of non purely
trigonal symmetry are present in the structure \cite{Muller88}:
Despite the $C_{3}$ point symmetry of the complex in the average
structure, disordered water molecules in the lattice can lead to
local non-trigonal components, thereby inducing isosceles
($J_{12} = J$ and $J_{13} = J_{23} = J'$) or scalene triangles
($J_{12} > J_{13} > J_{23}$ in eq.~\ref{eq:Heis}) owing to the
slight bond distance and angle differences. The GS degeneracy is
then lifted with a gap $2\Delta_{2}$ between the two Kramers
doublets given by $2\Delta_{2} = (u^2 + v^2 - uv)^{1/2}$ with $u=
J_{12} - J_{23}$ and $v= J_{13} - J_{23}$. No splitting of the
$S=3/2$ state is induced by this symmetry reduction. The lifting
of the ground states degeneracy corresponds to a partial removal
of the frustration in the equilateral triangular model.

In the isosceles case with $J > J'$ \cite{comment3}, the energy
gap between the two Kramers doublets becomes $2\Delta_{2} = J-J'$
where $\Psi^{\pm}_{0}$ is the ground state separated by
$2\Delta_{2}$ from the $\Psi^{\pm}_{1}$ state. Using this model
to analyze the INS data, we obtain $J = 0.223$ meV and $J'=
0.188$ meV, and the corresponding calculated $\chi T$ curve
between $100$ K and $1.8$ K is shown in fig.~\ref{fig:2} (dashed
curve). The result is not distinguishable to that of the
equilateral model in this temperature range since $T \geq |J-J'|$
down to $T=1.8$ K. It shows, in passing, that the powder
susceptibility alone, above $T=1.8$ K (i.e. $T \gg 2 \Delta$),
does not provide conclusive information on the important splitting
within the $S=1/2$ ground states \cite{Gatteschi96}. Both the DM
interaction and a symmetry reduction of the equilateral triangle
can induce a splitting in the ground state, and the observed $35
\un{\mu eV}$ gap in $\chem V_{15}$ may be a combination of both
processes.

\section{Conclusion}

Our neutron scattering results on a deuterated powder of $\chem
V_{15}$ show directly and unambiguously that the {\it effective}
Heisenberg coupling between the 3 S=1/2 spins in the triangle is
$J_{0} \approx 0.211$ meV and the presence of a splitting in the
S=1/2 ground states of $2\Delta \approx 35 \un{\mu eV}$. This
splitting is twice as large as that reported in
Ref.\cite{Chiorescu00} from a two-level Landau-Zener model
analysis of the magnetisation relaxation. Our value is more
accurate as it is based on a direct spectroscopic observation of
the gap. This splitting can be attributed to deviations from the
trigonal symmetry and/or Dzyaloshinskii-Moriya antisymmetric
interactions.

\acknowledgments This work has been supported by the Swiss
National Science Foundation and by the TMR program Molnanomag of
the European Union (No: HPRN-CT-1999-00012). One of us (A.M)
acknowledges the support of the Deutsche Forschungsgemeinschaft
and the Fonds der Chemischen Industrie.

{}

\pagebreak

\begin{thebibliography}{}
\bibitem{comment1} Within the limited $Q$-range available ($0.6 \leq Q \leq
1.5$ \AA\ $^{-1}$), we find that the $Q$-dependence of the
transition intensities agrees with the expected behaviour for a
polycrystalline trinuclear cluster, $I = I_{0}|F(Q)|^{2}(1 -
[\sin(QR_{0})/QR_{0}])$, where $|{F(Q)}|$ is the magnetic form
factor and $R_{0} = 5.96 $ \AA\ .
\bibitem{comment2} The INS intensities being controlled by the population factor of
the initial state, transitions within the S=3/2 states (if split)
would have a smaller intensity at $T=1.5$K than at $T=10$K.
\end{thebibliography}
\end{document}